\documentclass[showpacs,preprint,amsmath,superscriptaddress,prl]{revtex4}  % alvaro modified
\usepackage[dvips]{graphicx}
\usepackage{epsfig}
\usepackage{color}
\usepackage{subfigure}
\usepackage{soul}

\begin{document} 
\title{
Exact derivation of a finite-size-scaling law 
and corrections to scaling
in the geometric Galton-Watson process
} 
\author{\'Alvaro Corral}
% 
%\email{acorral@crm.es} 
\affiliation{Centre de Recerca Matem\`atica,
Edifici C, Campus Bellaterra,
E-08193 Barcelona, Spain.
} 
\affiliation{Departament de Matem\`atiques,
Facultat de Ci\`encies,
Universitat Aut\`onoma de Barcelona,
E-08193 Barcelona, Spain}
\author{Rosalba Garcia-Millan}
\affiliation{London Mathematical Laboratory, 14 Buckingham Street, London WC2N 6DF, UK}
\affiliation{Mathematical Institute, Andrew Wiles Building, Woodstock Road, Oxford OX2 6GG, UK}
\author{Francesc Font-Clos}
\affiliation{ISI Foundation, Via Alassio 11/c, 10126 Torino, Italy} 
\begin{abstract} 
The theory of finite-size scaling explains how the singular behavior 
of thermodynamic quantities in the critical point of a phase transition
emerges when the size of the system becomes infinite.
Usually, this theory is presented in a phenomenological way.
Here, we exactly demonstrate the existence of a finite-size scaling law for the 
Galton-Watson branching processes when the number of offsprings of each
individual follows either a geometric distribution or a generalized geometric distribution.
We also derive the corrections to scaling and the limits of validity of the finite-size scaling law
away the critical point.
A mapping between branching processes and random walks allows us to establish
that these results also hold for the latter case, for which the order parameter turns out to be the 
probability of hitting a distant boundary.
\end{abstract} 
% \pacs{64.60.av, 91.30.-f, 81.30.Kf,05.50.+q}

\maketitle

\section{Finite-size scaling}

Statistical mechanics provides 
a complete explanation of the thermodynamic (that is, macroscopic) 
properties of systems in terms of their microscopic laws when the
so-called thermodynamic limit is considered --
the limit of infinite system size 
\cite{Barber,Kuzemsky}.
However, there is a natural and increased interest in the properties
of small systems \cite{Bustamante,Parrondo}, 
i.e., systems whose size cannot be considered infinite.
What is finite and what is infinite is a relative matter, 
as systems displaying a continuous or second-order phase transition illustrate.
{The key issue is that the size of the system
needs to be measured in terms of its correlation length.}
For these systems
a useful tool to deal with finite-size effects near the critical point 
of the transition is finite-size scaling \cite{Brezin,Barber,Privman}.

Let us
consider a simple ferromagnetic system, 
whose thermodynamic variables are:
the magnetic moment per particle $\mu$ (proportional to magnetization),
the absolute temperature $T$,
and the magnetic field $H$.
It is convenient to rescale (and center) $T$ 
by means of the critical temperature $T_c$,
yielding the reduced temperature
$\tau =(T-T_c)/T_c$,
and to rescale $H$ by $k_B T$ (with $k_B$ Boltzmann constant),
yielding the reduced magnetic field $h=H/(k_B T)$.
Additionally, one may consider a system of units 
in which $\mu$ and $h$ are dimensionless.
The former, $\mu$, will be the order parameter, 
whereas $h$ and $\tau$ are control parameters.

``Near'' the critical point of the transition,
defined by $\tau=h=0$,
the equation of state fulfills a scaling law,
which gives $\mu$ as a function of $\tau$ and $h$ as
\begin{equation}
   \mu = {|h|^{\beta/\Delta}} \hat F_{\pm} \left(\frac \tau {|h|^{1/\Delta}}\right),
\label{scaling_muth}
\end{equation}
where $\beta$ and $\Delta$ are critical exponents, 
and $\hat F_\pm$ represents two scaling functions, 
one ($+$) for $h>0$ and another one $(-)$ for $h<0$.
The scaling law (\ref{scaling_muth}) indicates the invariance of the equation of state
under appropriate scale transformations (which are linear transformations
of the axes $\mu, \tau$ and $h$). 
{By the universality property many different systems share the same
values of the critical exponents and the same scaling functions,
and then the scaling law (\ref{scaling_muth}) constitutes a law
of corresponding states \cite{Stanley}.}

For instance, for the mean-field theory or the Landau theory of the Ising model \cite{Yeomans1992,Christensen_Moloney},
$\beta=1/2$, $\Delta=3/2$, and the scaling function $\hat F_\pm$ is given
by the two real solutions of $x=|\hat F_{\pm}(x)|^{-1} - |\hat F_{\pm}(x)|^2/3$, 
Ref. \cite{Wolfram_Vieta}.
This yields 
\begin{equation}
\hat F_\pm(x) = \left\{ \begin{array}{ll}
\pm \sqrt{-3x}  & \mbox{ for } x\rightarrow -\infty\\
\pm \sqrt[3]{3} & \mbox{ for } x=0\\
\pm 1/x & \mbox{ for } x\rightarrow \infty\\
\end{array}\right.
\end{equation}
and substituting into the scaling law (\ref{scaling_muth}),
one gets
\begin{equation}
\mu = \left\{ \begin{array}{ll}
\pm \sqrt{-3 \tau}  & \mbox{ for } \tau<0 \mbox{ and } h\rightarrow 0\\
\sqrt[3]{3h} & \mbox{ for } \tau=0 \\
h/\tau & \mbox{ for } \tau>0 \mbox{ and } h\rightarrow 0\\
\end{array}\right.
\label{threelaws}
\end{equation}
leading to the equation of the spontaneous magnetization, the critical-isotherm equation, 
and the Curie-Weiss law, respectively \cite{Yeomans1992}.
{As $\hat F_\pm(x)$ is a smooth function, 
it is only at the critical point that a sharp transition emerges.}

It is important that
the correlation length $\xi$ fulfills a scaling law analogous to Eq. (\ref{scaling_muth}),
\begin{equation}
 \xi=  {|h|^{-\nu/\Delta}} \hat G_{\pm} \left(\frac \tau {|h|^{1/\Delta}}\right),
\label{xi}
\end{equation}
with $\nu$ another critical exponent
and $\hat G_\pm$ another pair of scaling functions.
Then the main fact of critical phenomena is that $\xi$ diverges (goes to $\infty$) 
right at the critical point (as $\nu$ and $\Delta$ are positive).
For instance, at the critical isotherm, $\tau=0$, 
one has $\xi \propto 1/|h|^{\nu/\Delta}$, whereas
at zero field, $\xi \propto 1/|\tau|^\nu$.

Strictly, all these equations are 
only valid in the thermodynamic limit.
For a system of finite size $L$ {(in all dimensions \cite{Barber})}
the correlation length cannot be infinite.
When $L$ is much larger than the correlation length one
does not expect that the finiteness of the system has any influence
on the behavior of the system; however, this is not the case
when $L$ becomes smaller than the correlation
length of the corresponding infinite system \cite{Barber}.
So, one can introduce a phenomenological additional dependence on $\xi/L$ 
in the equation of state \cite{Privman},
as
$
\mu = {|h|^{\beta/\Delta}} \hat F_{\pm} \left( \tau / {|h|^{1/\Delta}},\xi/L\right),
$
which, substituting the equation for $\xi$ [Eq. (\ref{xi})], can be written as
$\mu = |h|^{\beta/\Delta} \tilde F_{\pm} \left( \tau / {|h|^{1/\Delta}},
L|h|^{\nu/\Delta}\right),
$
or, equivalently,
\begin{equation}
\mu={L^{-\beta/\nu}} F
\left(
L^{1/\nu} \tau,
 L^{\Delta /\nu} h
\right),
\label{fssone}
\end{equation}
where the terms $|h|^{\beta/\Delta}$ and $ \tau / {|h|^{1/\Delta}}$ 
have been transformed to
$L^{-\beta/\nu}$ and $L^{1/\nu} \tau$, respectively.
The previous equation constitutes a finite-size scaling law or ansatz,
where now $\hat F_\pm$, $\tilde F_\pm$ and $F$
become bivariate scaling functions, 
with the latter unifying the positive and negative values of $h$.
The finite-size-scaling ansatz can be verified by
plotting $\mu L^{\beta/\nu}$ versus 
$\tau L^{1/\nu}$
and 
$h L^{\Delta /\nu}$;
if a data collapse emerges, this gives the shape
of the scaling function $F$.
{In this way, finite-size behavior is determined from the
critical exponents of the infinite system \cite{Barber}.}

Note that for a finite system with $h=0$ the system size $L$
plays a role similar to that of the inverse of the magnetic field in an infinite system, 
or more precisely, $L^{1/\nu}$ acts as $1/|h|^{1/\Delta}$,
and in this way, one expects that the first argument of the scaling function $F$ in
Eq. (\ref{fssone}) behaves, qualitatively, 
as the scaling function $\hat F_\pm$ in Eq. (\ref{scaling_muth}).
This implies that a sharp transition can only take place
for $L\rightarrow \infty$, i.e., 
in the thermodynamic limit.

\section{Phase transition in the Galton-Watson process}

The Galton-Watson process \cite{Galton_Watson,Harris_original} provides the simplest model for the growth 
(and degrowth) of a biological population \cite{branching_biology},
but it is equally applicable to the growth of a nuclear reaction \cite{neutron_fluctuations}, 
an earthquake \cite{Corral_FontClos}, 
or mean-field self-organized critical processes in general \cite{Corral_FontClos,Alstrom,Zapperi_branching,Pruessner_book}.
It belongs to a more general class of models known as branching processes.
The Galton-Watson process 
starts with one single element that replicates, producing more elements, 
called offsprings,
which also replicate, producing more elements and so on.
The model is stochastic, as the (total) number of offsprings produced by each element is random,
characterized by a distribution that is the same for all elements
and also independent of the number of offsprings of the other elements.

In mathematical terms, the probability that the number of offsprings $K$ of
one element takes the value $k$ is given by $P[K=k]$, with $k$ taking discrete values
from 0 to $\infty$.
In this paper we will consider that $P[K=k]$ is given by the geometric distribution, 
or by the generalized geometric distribution, but the model is totally general.
The distribution $P[K=k]$ completely defines the model, as, we insist, the number of offsprings of 
each element are identically distributed and independent.
The initial element defines the $0-$th generation, 
its offsprings are the first generation, and so on.
An index $t$ labels each generation.
The model does not incorporate time, 
but one can interpret $t$ as a discrete time.
An important auxiliary variable is $N_t$, which counts the number of elements
in each generation, starting with $N_0=1$ (one single original element).

The key question to ask is if the process gets extinct, i.e., $N_t=0$ at some $t\ge 1$, 
or not (where it goes on forever).
A fundamental result in the theory of branching processes 
\cite{Harris_original, Corral_FontClos} is that
the probability of extinction $P_{ext}$ can be obtained from
\begin{equation}
P_{ext} = 
\lim_{t\rightarrow \infty} P[N_t=0] = 
\lim_{t\rightarrow \infty} f^t(0),
\end{equation}
where $f^t(s)$ is the $t-$th composition
of the probability generating function $f(s)$ of the random variable $K$,
i.e., $f^t(s)= f(\dots f(f(s))\dots)$ (composed $t$ times), with 
\begin{equation}
f(s) =\sum_{k=0}^\infty P[K=k] s^k.
\end{equation}
As we iterate successive compositions of $f(s)$ starting from $s=0$,
the limit is given by the smallest fixed point $s^*$ of $f(s)$
in the interval $[0,1]$; so, $s^*$ necessarily satisfies $s^*=f(s^*)$,
but it is the smallest value in $[0,1]$ verifying such relation.

Introducing the probability of survival, or probability 
of non extinction $\rho$, fulfilling $P_{ext}=s^*=1-\rho$, 
the fixed-point condition becomes
\begin{equation}
1-\rho = 
\sum_{k=0}^\infty P[K=k] (1-\rho)^k.
\end{equation}
From here, it is clear by normalization that $\rho=0$ is a possible solution.
Expanding the equation up to second order in $\rho$ using the binomial theorem
one gets
\begin{equation}
1-\rho
\simeq
 \sum_{k=0}^\infty P[K=k] \left(1-k\rho + \frac{k(k-1)} 2\rho^2\right) 
\end{equation}
\begin{equation}
=1-\langle K \rangle \rho + \frac 1 2\langle K(K-1)\rangle \rho^2.
\end{equation}
The solutions, in terms of the mean number of offsprings, 
$m=\langle K \rangle$, and close to $m=1$, are then
\begin{equation}
\rho = 
\left\{
\begin{array}{ll}
0 & \mbox{ for } m \le 1, \\ 
2 \sigma_c^{-2} {(m-1)} &
\mbox{ for }  m \ge 1, \\  
\end{array}
\right.
\label{nosecual}
\end{equation}
where we have used that when $\rho$ is close to zero (from above)
$m$ is close to one, 
and therefore
$\langle K  ( K -1) \rangle=\sigma^2 + m (m-1) \simeq \sigma_c^2$,
where $\sigma_c^2$ is the variance of $K$ when its mean is one.
It can be proved that there are no other fixed points than the two above
\cite{Harris_original,Corral_FontClos}.

It is clear that the case in which the offspring distribution verifies $m=1$
is critical, in the sense that it separates two very different ``phases'' of the system:
extinction for sure if $m \le 1$ and non-sure extinction (and the possibility of a
``demographic'' explosion) for $m>1$.
Even more, 
this phase diagram 
is analogous to the spontaneous (zero-field) behavior of a magnetic system, 
Eq. (\ref{threelaws}),
if we identify $m-1$ with the control parameter $\tau$ and $\rho$ with the order parameter $\mu$,
and so we can talk about a phase transition in the Galton-Watson model
\cite{Corral_FontClos}
with critical point at $m=m_c=1$.
Note then that $\sigma_c^2$ becomes the variance of the number of offsprings
in the critical case.
There are, though, two quantitative differences:
$\beta=1$ (in contrast to $\beta=1/2$ in the magnetic example above)
and that the ordered phase (non-zero order parameter)
is above the critical point now.
Equation (\ref{nosecual}) also tell us that 
when the distance to the critical point, $m-1$, 
is rescaled by $\sigma_c^2$ 
the behavior of the transition is universal, i.e., 
independent on the underlying distribution of the number of offsprings $K$.

In this paper we investigate this phase transition 
for a finite number of generations, i.e., 
when the number of generations is limited by $t\le L$.
In a previous paper \cite{GarciaMillan} we expanded $f(f^t(0))$ around
the critical point $s^*$ to obtain a general finite-size-scaling law
for the probability of survival $\rho$.
Here we follow a different, more direct approach, 
particularized for a geometric distribution in the number of offsprings,
which will allow us to obtain also the corrections to scaling.

After the introduction to finite-size scaling in critical phenomena in the previous section
and the introduction to branching processes in this section,
in Sec. 3 we analyze the finite-size effects in the critical properties of 
the Galton-Watson process when the offspring distribution is given
by the geometric distribution.
Two different order parameters are explored, 
[$\rho$ and $\rho/(1-\rho)$],
and the corrections to scaling 
and the range of validitity of the scaling law are obtained as well.
We generalize the finite-size scaling law for the so-called 
generalized geometric distribution in Sec. 5.
Previously, in Sec. 4, we establish that our scaling law also describes 
the {escape} probability of a simple one-dimensional random walk.
An appendix gives some details of the calculations of Secs. 3 and 5.

\section{Finite-size scaling in the geometric Galton-Watson process}

We consider the Galton-Watson model with a finite number of generations $L$,
which means that the process is stopped when it reaches the $L-$th generation,
i.e., the elements of this generation are not allowed to replicate.
Viewing the process as a branched tree, $L$ becomes the 
height of the tree
and is therefore a measure of system size
(more precisely, the height of the tree is $L+1$, counting the $0-$th generation).

The
extinction of this process is given by the event $N_L=0$,
as extinction at any generation $t <L$ is included in the case $N_L=0$
(extinction is forever, as it is an absorbing state). 
In the same way as for an unbounded system, 
the probability of extinction will be 
\begin{equation}
P_{ext}(L) = 
P[N_L=0] = 
f^L(0)
\end{equation}
(we only make explicit the dependence on $L$,
but a hidden dependence exists in the parameters of the distribution of $K$,
in particular on $m$).
The probability of extinction is obtained then as the $L-$th composition 
of the probability generating function of the distribution of the number of offsprings,
but note that as $L$ is not infinite, $f^L(0)$ will not reach the fixed point $s^*$.
Although formally the problem is solved by the calculation of $f^L(0)$, 
in general it is not feasible to
arrive to an explicit expression for the composition, 
even for small values of $L$.

A remarkable exception is the case when $K$ follows the geometric distribution,
given by
\begin{equation}
P[K=k] = p q^k,
\label{geodist}
\end{equation}
for $k=0,1,\dots \infty$ (and zero otherwise) and with $q=1-p$.
The only parameter of the distribution is $p$, which is called the success
probability.
The geometric distribution has a straightforward interpretation in terms of
biological populations.
For instance, consider that the elements that replicate are female individuals,
and each female has a probability $q$ to produce another female
and a probability $p$ of producing a male.
Each female reproduces until it gets a male, 
and when the male is obtained the mother does not reproduce anymore.
Although getting a male is considered a ``success'' (this is just a name),
it is the female individuals what are counted as offsprings,
so $K$ counts the number of females disregading the male.
Note that another variant of the geometric distribution counts also the male, 
this would be for us a shifted geometric distribution and is not considered here.

The probability generating function of the geometric distribution turns out to be
\begin{equation}
f(s)=\sum_{k=0}^\infty p q^k s^k =\frac p {1-qs},
\label{fgeo}
\end{equation}
from which the mean is obtained as $m=\langle K \rangle = f'(1)=q/p$
and the variance as $\sigma^2=f''(1)-m(m-1)=q/p^2$,
see Ref. \cite{Corral_FontClos}.
Note that the critical point of the corresponding Galton-Watson process is
at $m=q/p=1$ and so $p_c=q_c=1/2$, with a critical variance $\sigma_c^2=2$.

The fundamental property (for our problem) of the geometric distribution 
comes from the fact that its probability generating function is a fractional linear function \cite{Harris_original},
also called a linear fractional function \cite{Karlin_Taylor}. 
In this case the successive compositions of $f(s)$ 
can be computed for any $L$,
yielding
\begin{equation}
f^L(s)=\frac{s_0-\kappa^L (s-s_0)/(s-1) }{ 1-\kappa^L (s-s_0)/(s-1) },
\end{equation}
see Ref. \cite{Karlin_Taylor} or Eq. (\ref{ladelapendice}) at our Appendix.
The constant $s_0$ is a fixed point of $f(s)$
different from 1 (this fixed point, $s_0$, always exists except for $m=1$), 
and the constant $\kappa$ is given in the Appendix.
Then, the probability of survival will be
\begin{equation}
\rho(L)=1-f^L(0)=\frac{1-s_0}{1-\kappa^L s_0},
\label{labuena}
\end{equation}
which contains the solution to our problem.

For the geometric distribution the fixed point $s_0$ is at
$s_0=p/q=m^{-1}$, and then $\kappa=p/q=m^{-1}$
(see Appendix); therefore, 
substituting into Eq. (\ref{labuena}) we get
\begin{equation}
\rho(L)=\frac{m^L(m-1)}{m^{L+1}-1}.
\label{scaling_geo}
\end{equation}
This exact equation provides the order parameter $\rho$
as a function of the control parameter $m$ for any system size $L$
(in the case of the geometric distribution).

In order to verify if a scaling law is fulfilled it is convenient 
to introduce the rescaled distance to the critical point, 
\begin{equation}
x=L^{1/\nu} (m-1),
\end{equation}
where the ``distance'' $m-1$ is rescaled (divided) by the term $1/L^{1/\nu}$,
with the value of the exponent $\nu$ unknown.
Substituting $m-1=x/L^{1/\nu}$ and 
\begin{equation}
m^L=\left(1+\frac x {L^{1/\nu}}\right)^L
\end{equation}
into Eq. (\ref{scaling_geo}) we observe that the rescaled survival probability 
$L^{1/\nu} \rho(L)$ in the limit $L\rightarrow \infty$
either tends to zero or infinite (depending on the sign of $x$
and on whether $\nu>0$ or $\nu <0$), except in the case $\nu=1$.
For $\nu=1$ and close to the critical point,
the limit of $L^{1/\nu} \rho(L)$ is a positive value
that only depends on $x$, which is the signature of a scaling law,
\begin{equation}
L \rho(L) \propto {F}(x),
\end{equation}
with ${F}$ the scaling function.

Indeed, rewritting Eq. (\ref{scaling_geo}) in terms of $x$, using that
$m^L\rightarrow e^x$ for $\nu=1$ leads to
\begin{equation}
\rho(L)\simeq \frac{e^x x/L}{e^x-1},
\end{equation}
up to the lowest order in $L^{-1}$. 
Taking into account that the variance at the critical point is $\sigma_c^2=2$,
the scaling law can be written as
\begin{equation}
\rho(L)\simeq \frac 1 {\sigma_c^2 L}\left(\frac{2x e^x }{e^x-1}\right)=
\frac 1 {\sigma_c^2 L}{F}(x),
\label{scalingdeverdad}
\end{equation}
with scaling function
\begin{equation}
{F}(x)=\frac{2x e^x }{e^x-1},
\label{scalingfunction}
\end{equation}
in total agreement with Ref. \cite{GarciaMillan}.
The reason to introduce the value of $\sigma_c^2$ 
will become more clear when we consider the generalized geometric case, 
in Sec. 5.

It is important that
the scaling function (\ref{scalingfunction}) fulfills
\begin{equation}
F(x) = \left\{
\begin{array}{ll}
-2xe^x & \mbox{ for } x \rightarrow -\infty,\\
2 & \mbox{ for } x=0,\\
2x & \mbox{ for } x \rightarrow \infty.\\
\end{array}
\right.
\end{equation}
{Although our calculation does not include the critical case, 
$x=0$, the Appendix shows that indeed the critical case
is also described by the value of the scaling function $F$ 
at $x=0$.
{Therefore, there is a removable singularity at $x=0$.}
The limit behavior of $F$,} 
substituted into the scaling law, leads to
\begin{equation}
\rho(L) = \left\{
\begin{array}{ll}
2 \sigma_c^{-2} (1-m) e^{-L(1-m)} & \mbox{ for } m < 1 \mbox { and } L \rightarrow \infty,\\
2 \sigma_c^{-2} L^{-1} & \mbox{ for } m=1, \\
%%\mbox { and } L \rightarrow \infty\\
2 \sigma_c^{-2} (m-1) & \mbox{ for }  m > 1 \mbox { and } L \rightarrow \infty.\\
\end{array}
\right.
\end{equation}
We see that the infinite-size case, Eq. (\ref{nosecual}),
is recovered when $L$ is infinite,
and that it is only in this case that a sharp transition exists.

Comparison with Eq. (\ref{threelaws}) allows one to see which is the equivalent
of the 
``critical isotherm'' and
``spontaneous magnetization'' laws
for the Galton-Watson process.
For the latter case we see that $\beta=1$.
The Curie-Weiss law is not fulfilled as $\rho$ does not decay as a power law
in $L$ but exponentially for $m< 1$.

\begin{figure*}
\begin{center}
\includegraphics[width=10cm]{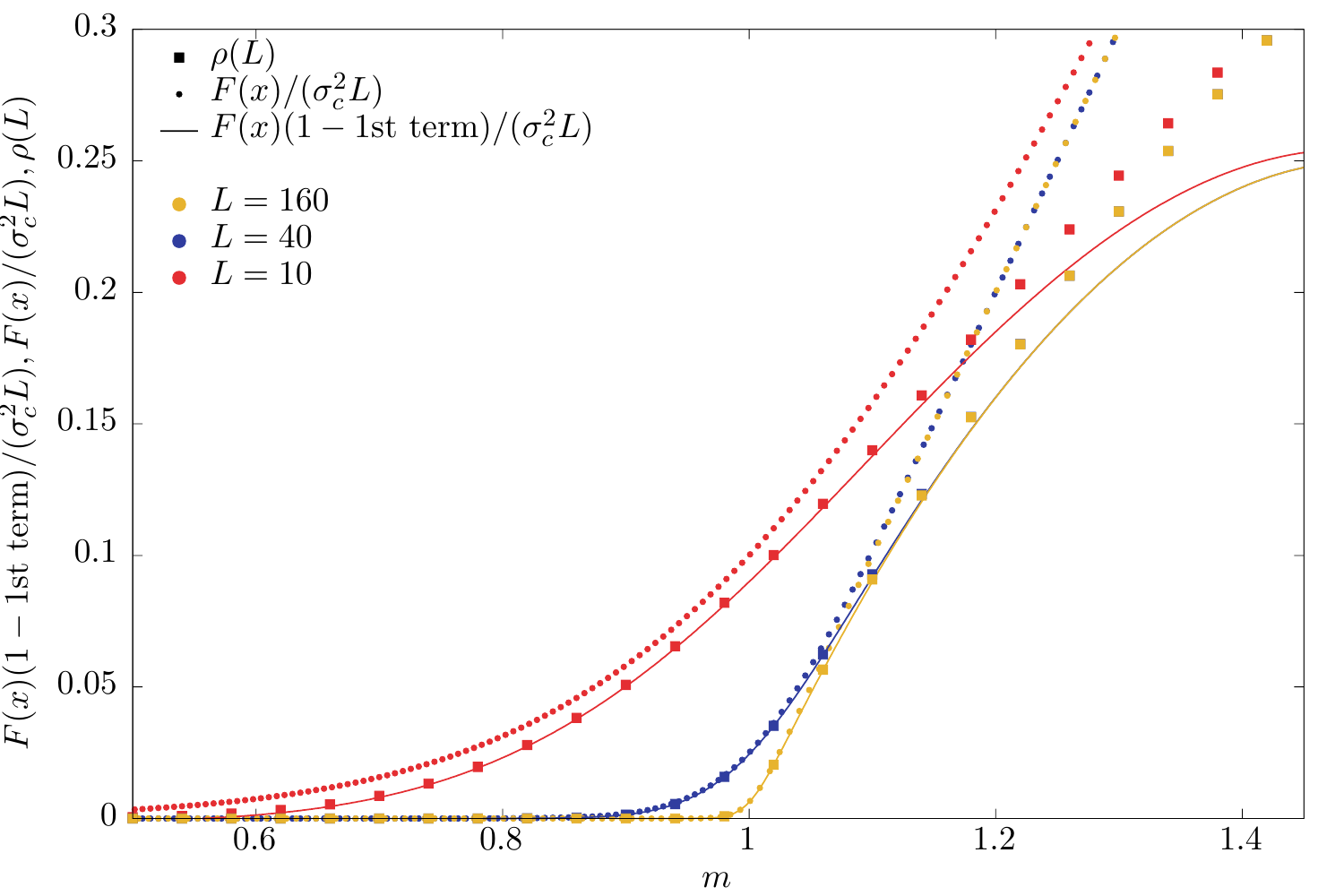}
\includegraphics[width=10cm]{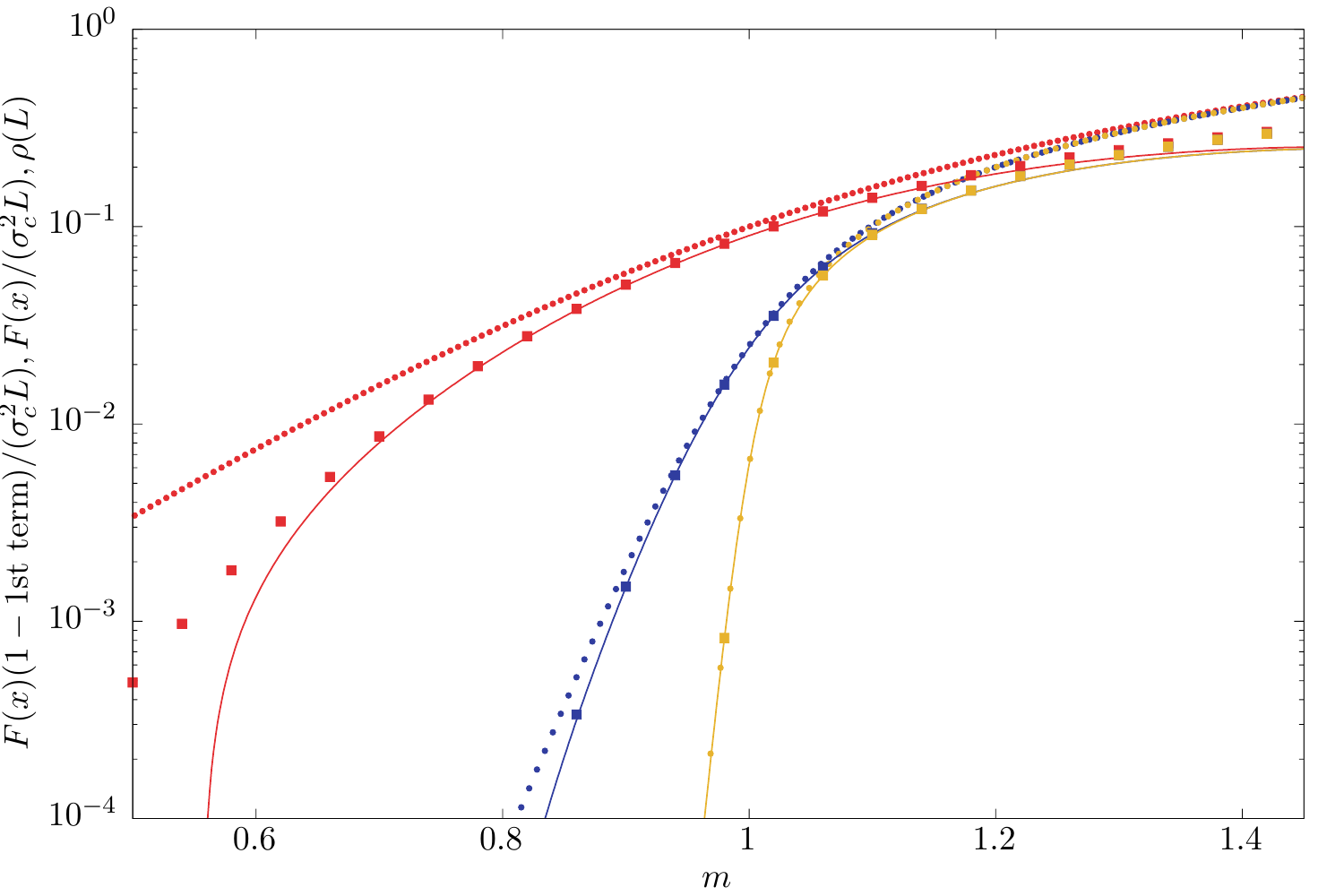}
\includegraphics[width=10cm]{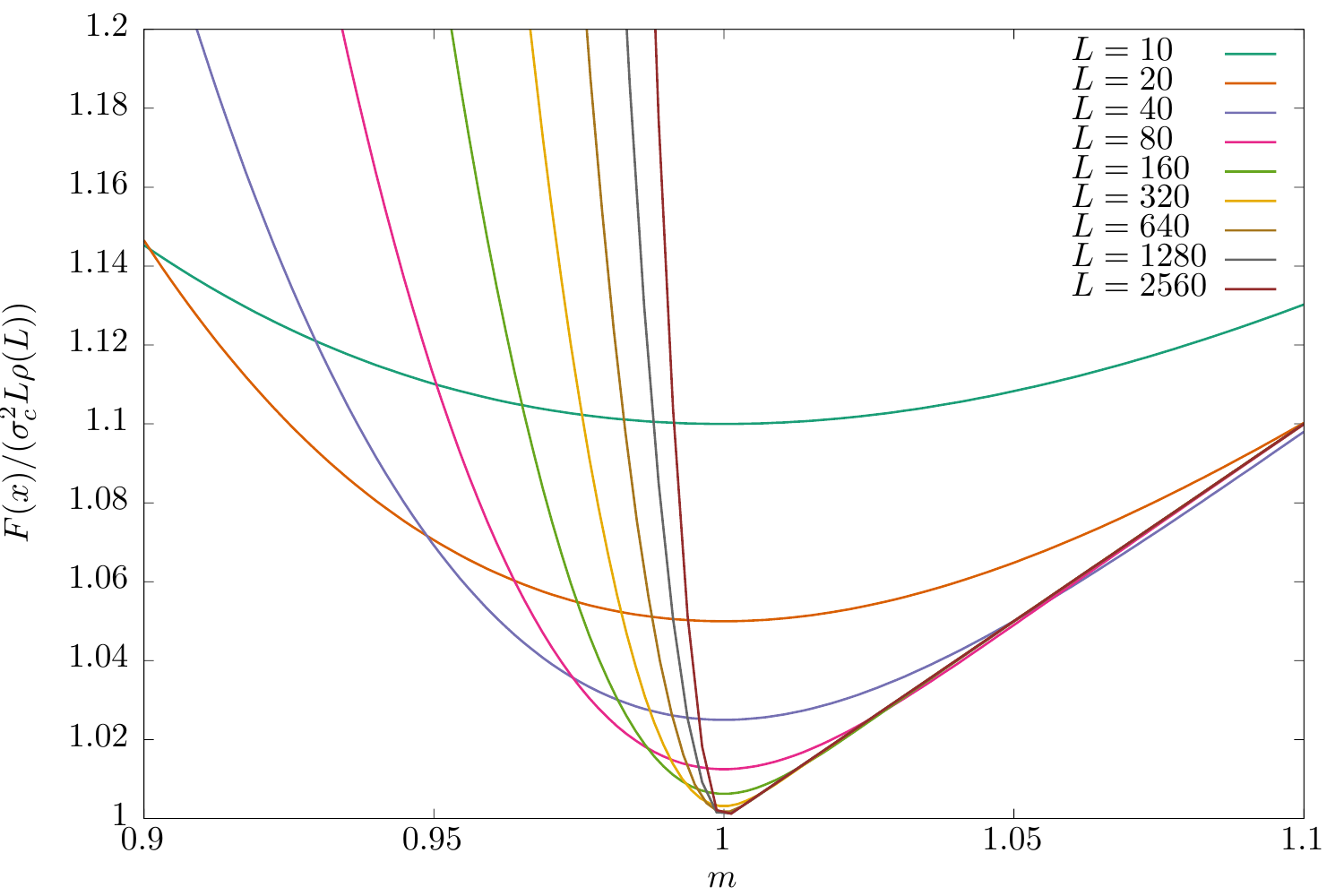}
\end{center}
\caption{
(a) Comparison of the exact probability of survival, $\rho(L)$, given
by Eq. (\ref{scaling_geo}), with the approximations given by the scaling law
Eq. (\ref{scalingdeverdad}) and by the scaling law with the first correction to scaling,
Eq. (\ref{scalingpluscorrection}), 
for different $m$ and $L$.
(b) The same taking the $y-$axis logarithmic.
(c) The same data, taking the ratio 
between the approximation given by the scaling law [$F(x)/(\sigma_c^2 L)$],
Eq. (\ref{scalingdeverdad}), and
the exact value of $\rho(L)$.
Larger values of $L$ are included in this case.
\label{figerror}
}
\end{figure*}

We may also obtain the corrections to scaling,
taking care of terms beyond the leading one.
Going back to Eq. (\ref{scaling_geo}),
we substitute there the exact expression
$m^L=(1+x/L)^L=e^x(1 +\sum_n a_n)$,
with $a_1=-x^2/(2L)$, $a_2=x^3/(3L^2)$, etc.,
then,
\begin{equation}
\rho(L) =
\frac{ e^x(1+\sum a_n) x/L}{e^x (1+\sum a_n)(1+x/L)-1}
= \frac{2xe^x}{\sigma_c^2 L (e^x-1)}
\left( \frac{1+\sum a_n} {1+u \sum b_n}\right)
\end{equation}
\begin{equation}
= \frac{F(x)}{\sigma_c^2 L}
\left(1+\sum a_n\right)\left[1-u \sum b_n+u^2 \left(\sum b_n
\right)^2 + \dots \right]
\end{equation}
\begin{equation}
= \frac{F(x)}{\sigma_c^2 L}
\left[1+\sum a_n -u \sum b_n - u\left(\sum a_n\right)\left(\sum b_n\right) + 
u^2\left(\sum b_n\right)^2 + \dots \right],
\end{equation}
with $u=e^x/(e^x-1)$ and $\sum b_n = x/L + (1+x/L) \sum a_n$.
The first terms of the different sums are
\begin{equation}
\sum a_n = -\frac{x^2}{2L} + \frac{x^3}{3L^2}  + \frac{x^4}{8L^2} -\frac{x^4}{4L^3}+ \dots
\end{equation}
\begin{equation}
\sum b_n = \frac x L -\frac{x^2}{2L} -\frac{x^3}{6L^2}+\frac{x^4}{8L^2} + \frac{x^4}{12L^3}+ \dots
\end{equation}
\begin{equation}
\left(\sum b_n \right)^2= \frac {x^2}{L^2} -\frac{x^3}{L^2} + \frac{x^4}{4L^2}-\frac{x^4}{3L^3}+\dots
\end{equation}
\begin{equation}
\left(\sum a_n \right)\left(\sum b_n \right)= -\frac {x^3}{2L^2} +\frac{x^4}{4 L^2} +\frac{x^4}{3L^3}+ \dots
\end{equation}
\begin{equation}
\left(\sum b_n \right)^3= \frac {x^3}{L^3} -\frac{3 x^4}{2L^3}+\dots
\end{equation}
\begin{equation}
\left(\sum a_n \right)\left(\sum b_n \right)^2=  -\frac{x^4}{2L^3}+ \dots
\end{equation}
Let us study the behavior as far from the critical point as possible.
Below it ($x<0$), we take
$x \rightarrow -\infty$ and then $u\rightarrow 0$ (exponentially in $x$); 
therefore, only $\sum a_n$ contributes and 
we get
\begin{equation}
\rho(L)=\frac{F(x)}{\sigma_c^2 L} \left(1-\frac{x^2}{2L}+\dots\right)
\label{suspiro}
\end{equation}
so, the first correction-to-scaling term goes as $-x^2/(2L) = -L(m-1)^2/2$.
This means that if this term is of order $\varepsilon$
(i.e., $L(m-1)^2/2=\varepsilon$)
all other terms are of higher order in $\varepsilon$,
in the limit $L\rightarrow \infty$.
This is so because the rest of terms are of the form
\begin{equation}
\frac{x^{2k-1}}{L^k}, \frac{x^{2k-1}}{L^{k+1}}, \dots \frac{x^{2k-1}}{L^{2k-2}},  
\end{equation}
and 
\begin{equation}
\frac{x^{2k}}{L^k}, \frac{x^{2k}}{L^{k+1}}, \dots \frac{x^{2k}}{L^{2k-1}}.
\end{equation}
Above the critical point ($x>0$) we consider
$x \rightarrow \infty$, then, $u\rightarrow 1$ and 
the sums lead to the cancellation of all terms that are not powers of $x/L$,
so 
\begin{equation}
\rho(L) =
\frac{F(x)}{\sigma_c^2 L}\left(1
-\frac x L + \frac{x^2}{L^2} - \frac{x^3}{L^3} + \dots \right).
\label{massuspiro}
\end{equation}
The first correction to scaling is given by the term $-x/L$.
If we impose this to be of order $\varepsilon$, (i.e.
$\varepsilon=x/L=m-1$), we will obtain the limit of
validity of the scaling law above the critical point.
In summary, the scaling law will hold in the range
\begin{equation}
1-\sqrt{\frac{2\varepsilon}L} <
m < 1+\varepsilon
\label{wepropose}
\end{equation}
with $\varepsilon \ll 1$.
For instance, for a 5 \% error
[defined as the ratio between the approximation given by the scaling law 
and the exact $\rho(L)$, Eq. (\ref{scaling_geo})],
$\varepsilon =0.05$ and then
$1-\sqrt{0.1/L} < m < 1.05$. 
Figure \ref{figerror} shows that this is valid for $L-$values above 40 
for $m<1$ and above 160 for $m>1$.
Note that the range of validity that we obtain, Eq. (\ref{wepropose}), 
is much larger than the one implicit in Ref. \cite{GarciaMillan},
$1-c/L < m  <1+c/L$, with $c$ a constant.
If we do not take the limits $x\rightarrow \pm \infty$, we have, 
keeping terms up to first order in $1/L$,
\begin{equation}
\rho(L)
=\frac{F(x)}{\sigma_c^2 L}\left(1-\frac{2xe^x-x^2}{2L(e^x-1)}+\dots\right),
\label{scalingpluscorrection}
\end{equation}
which is also shown in Figs.  \ref{figerror} a and b.

\begin{figure*}
\begin{center}
\includegraphics[width=15cm]{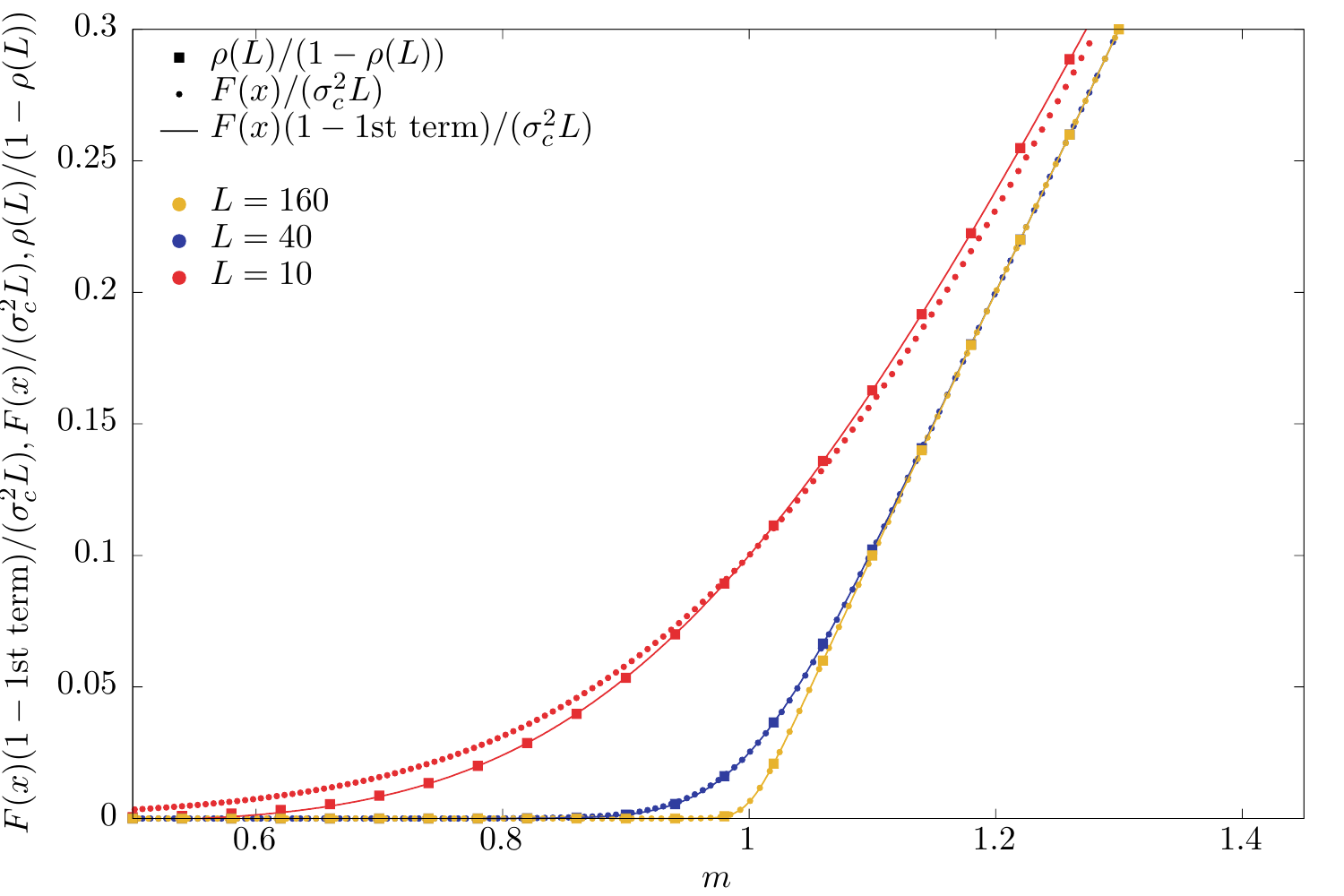}
\end{center}
\caption{Same as Fig. \ref{figerror}a, but replacing the order parameter $\rho(L)$ by $\rho(L)/[1-\rho(L)]$.
The exact behavior is given by Eq. (\ref{nuevaultima}),
and the scaling law with the first correction to scaling is given
by Eq. (\ref{nuevasl}).
It becomes clear how
the performance of the finite-size scaling law is even better than for $\rho(L)$, in particular
for $m>1$.
\label{dos}
}
\end{figure*}

A scaling law with a broader range of validity is obtained
taking as an order parameter not $\rho$ but $\rho/(1-\rho)$.
This is just the ratio between the number of realizations that survive
at $t=L$ and the number that are extinct at $t=L$.
From Eq. (\ref{scaling_geo}) we obtain 
\begin{equation}
\frac{\rho(L)}{1-\rho(L)} = \frac{m^L(m-1)}{m^L-1},
\label{nuevaultima}
\end{equation}
and proceeding as in the preceding case, we get
\begin{equation}
\frac{\rho(L)}{1-\rho(L)} =
\frac{ e^x(1+\sum a_n) x/L}{e^x (1+\sum a_n)-1}
= \frac{2xe^x}{\sigma_c^2 L (e^x-1)}
\left( \frac{1+\sum a_n} {1+u \sum a_n}\right)
\end{equation}
\begin{equation}
= \frac{F(x)}{\sigma_c^2 L}
\left[1+(1-u)\sum a_n - u(1-u)\left(\sum a_n\right)^2 
 + u^2(1-u)\left(\sum a_n\right)^3 + \dots \right].
\end{equation}
The factors $u^k(1-u)=-e^{kx}/(e^x-1)^{k+1}$ 
go to zero exponentially fast when $x\rightarrow \pm \infty$, 
except the first one ($k=0$) when $x\rightarrow -\infty$, 
for which $u\rightarrow 1$. 
This is the only contribution away from the critical point, 
and so (below the critical point) the correction to scaling goes as $-x^2/(2L)$.
The range of validity of the scaling law is then given by
\begin{equation}
m > 1- \sqrt{\frac {2\varepsilon} L},
\label{largerrange}
\end{equation}
i.e., the scaling law is valid arbitrarily far from the fixed point
in the supercritical region, as the correction term there decays
exponentially fast in $x$.
If we keep $x$ finite and terms up to first order in $1/L$ 
we arrive at
\begin{equation}
\frac{\rho(L)}{1-\rho(L)} = \frac {F(x)}{\sigma_c^2 L} 
\left(1+\frac{x^2}{2 L (e^x-1)}+\dots \right).
\label{nuevasl}
\end{equation}
This can be verified in Fig. \ref{dos}, where the scaling law 
describes system sizes as small as $L=10$ arbitrarily far from the critical point in the supercritical region.

\section{Applicability to random walks}

Thanks to a well-known mapping between branching processes
and random walks \cite{Harris52}, 
our finite-size scaling law is also applicable to the latter system.
In concrete, a one-dimensional random walk can be obtained
from the geometric Galton-Watson branching process
by following the branches sequentially.
Instead of considering that each generation $t$
of the process is generated in parallel from the previous one
(as the identification of the index $t$ with time suggests)
one changes the order in which offsprings appear.
The position of a walker in the tree associated to the branching
process determines which element (which node of the tree)
replicates.

The walker is initially located at
 the root (the element at the $0-$th generation),
and moves to one of the elements in the first generation
(it does not matter which one).
If this element has its own offsprings,
the walker moves to one of this, and so on. 
A branch is followed sequentially until the branch gets extinct
(the last element has no offsprings), and then the walker moves back
to the parent  of the last element
(from generation $t$ to $t-1$);
if this parent has more offsprings the walker follows
the branch of one of the remaining offsprings;
if not, the walker moves back to the 
previous parent (at generation $t-2$) and so on.
Note then that the walker passes twice through each link or edge between parent and offspring.
If, arbitrarly, we consider that the root is at the bottom of the tree
(as in real, biological trees!) and each new generation is one level above the
previous one, the walker travels up and down through all the tree.

The one-dimensional random walk is obtained from the projection
of the position of the walker on the axis counting the number of generations,
so, the $t-$axis of the branching process becomes the spatial axis
of the random walk.
Then, the walker moves up with probability $q$ and down with probability $p$
(the parameters of the geometrical distribution).
Notice that the mapping is possible and exact because the number of offprings follows the geometric distribution, 
Eq. (\ref{geodist}).

The finite-size condition imposed to the branching process translates 
into the existence of a reflecting boundary at $t=L$ for the random walk,
and then, the probability of survival $\rho$ of the branching process
turns out to be the probability of hitting the reflecting boundary, $P_{hit}$,
for the random walk. 
This also has an absorbing boundary at $t=-1$, where the walk dies
(after a duration equal to twice the number of elements, minus one).

After all these considerations, the mapping is established, and
we can write a finite-size scaling relation 
for the hitting probability,
\begin{equation}
P_{hit}(L) \simeq \frac 1 {2L} {F(x)} 
\end{equation}
with $F(x)$ given by Eq. (\ref{scalingfunction})
and
\begin{equation}x=L(m-1) \simeq 2L \left(q-\frac 1 2\right).\end{equation}
Remember that this is
valid for large $L$ and close to the critical point
$q=q_c=1/2$.
In particular, the corrections to scaling of the previous section
also hold when the relationships are written in terms of $m$
or $x=L(m-1)$.

{In fact, the previous scaling law describes the
probability that a random walk starting next to the absorbing boundary
hits the other boundary,
independently of the nature of the latter (reflecting or not),
as it is only the first-passage time what matters.}
In this way, the one-dimensional random walk, 
the simplest system in statistical physics,
displays a continuous phase transition with finite size scaling,
for which the corrections to scaling can be easily obtained as well.

\section{The generalized geometric distribution}

The previous analysis of the geometric Galton-Watson process in terms
of fractional linear functions (see Appendix) 
suggests a generalization of the problem.
We may consider the generalized geometric distribution, in which the zero-offspring
probability, $P[K=0]$, is released from following the geometric distribution and instead
it takes a free value $p_0$, which is a new parameter. 
The rest of values of $K$ follow the geometric distribution, but rescaled by $(1-p)/(1-p_0)$
(because of normalization).
In a formula,
\begin{equation}
P[K=k] = 
\left\{
\begin{array}{ll}
p_0  & \mbox{ for } k=0\\
(1-p_0) p q^{k-1} & \mbox{ for } k=1,2, \dots\\
\end{array}
\right.
\end{equation}
and zero otherwise.
We recover the usual geometric distribution for $p_0=p$.
The generating function is indeed a fractional linear function, 
\begin{equation}
f(s) = \frac{p_0 + (p-p_0) s}{1-qs}, 
\label{ggeopgf}
\end{equation}
which yieds $m=f'(1)=(1-p_0)/p$ and 
$\sigma^2=(1+p_0-p)(1-p_0)/p^2$.
The critical point turns out to be at $p_c=(1-p_0)$

The analysis of Sec. 3 is fully applicable in this case, 
in particular Eq. (\ref{labuena}).
We need to know that
$s_0=p_0/q$
and $\kappa=m^{-1}$ (see Appendix);
in fact, we write $s_0$ as a function of $m$ and $p_0$, 
which is $s_0=p_0 m /(m-q_0)$, with $q_0=1-p_0$.
Notice that we study the transition keeping fixed $p_0$.
Substituting into the formula for the order parameter $\rho(L)$, 
Eq. (\ref{labuena}), we arrive at
\begin{equation}
\rho(L)=
\frac{m^L(1-s_0)}{m^L-s_0}=
\frac{m^L (m-q_0-p_0 m)}{m^{L}(m-q_0)-p_0 m}
\end{equation}
\begin{equation}
=
\frac{m^L(m-1)(1-p_0)}{m^{L}(m-1+p_0)-p_0 m}.
\label{ocho}
\end{equation}
Introducing again the rescaled variable $x=L^{1/\nu} (m-1)$,
and taking the limit $L\rightarrow \infty$, the only non trivial
limit arises for $\nu=1$. 
In this case, up to first order in $1/L$ and introducing the critical variance
$\sigma_c^2=2p_0/(1-p_0)$, we get
\begin{equation}
\rho(L) \simeq \left(\frac{1-p_0}{2p_0}\right) \frac 1 L \left(\frac {2xe^x}{e^x-1}\right)
=\frac 1 {\sigma_c^2 L} F(x),
\end{equation}
which is the same scaling law as for the geometric case, 
with the scaling function $F(x)$ given again by Eq. (\ref{scalingfunction}).

\section{Summary}

We have presented here direct analogies between branching processes and 
thermodynamic phase transitions.
We have considered the classical Galton-Watson model of branching processes
when the number of offsprings $K$ per element is given by the geometric distribution.
This process has as natural control and order parameters 
the mean value of $K$ and the probability of survival $\rho$, respectively.
We study finite-size effects by imposing an upper limit $L$ to the number of generations.
After obtaining the exact expression for the equation of state, that is, 
the dependence of the order parameter with the control parameter, Eq. (\ref{scaling_geo}), 
we introduce the rescaled distance
to the critical point, $x=L^{1/\nu}(m-1)$.
When $\nu=1$ we demonstrate that a finite-size scaling law, 
Eq. (\ref{scalingdeverdad}), emerges in the limit $L\rightarrow \infty$.

In general, the theory of critical phenomena does
``not explain why in some systems
scaling holds for only 1-2 \% away from the critical
point and in other systems it holds for 30-40 \% away''
\cite{Stanley_rmp}.
In particular, finite-size scaling should work 
when the system size tends to infinite 
and the control parameter approaches the critical point;
nevertheless, in practice, finite-size scaling predictions turn out to apply
to rather small systems at a non-negligible distance from the critical point \cite{Barber}.
We provide a quantitative derivation of these limits for the finite-size scaling behavior of the Galton-Watson process,
Eq. (\ref{wepropose}), thanks to the calculation of the corrections to scaling, 
Eqs. (\ref{suspiro}) and (\ref{massuspiro}), or Eq. (\ref{scalingpluscorrection}).
If we define an alternative order parameter as $\rho/(1-\rho)$, the same scaling law holds, 
but with a larger range of validity, given by Eq. (\ref{largerrange}).
In this case the corrections to scaling are given by Eq. (\ref{suspiro}), below the critical point or
by Eq. (\ref{nuevasl}), in general.

A straightforward mapping between branching processes and random walks
allows one to establish that 
all our results for the survival probability of a geometric Galton-Watson process are equally valid
for the probability that a one-dimensional random walk, starting above but close to an absorbing origin
and evolving through $\pm 1$ increments, reaches a distance to the origin equal to $L$. 
In this way, a subcritical Galton-Watson process corresponds to a random walk with a bias to the negative ($-1$)
increment, for which the hitting probability becomes zero as $L\rightarrow \infty$.
On the other hand, the supercritical case corresponds to a random walk with a positive bias in the increment,
for which there exists a non-zero probability that never returns to the origin in the limit $L\rightarrow \infty$.
Obviously then, the critical case is the one of a fair random walk.
To the best of our knowledge, the one-dimensional random walk provides the simplest
example of a system exhibiting a finite-size scaling law.  
Therefore, the analogies between branching processes and equilibrium phase transitions
are totally applicable to the one-dimensional random walk.

\section{Acknowledgements}

{G. Pruessner encouraged us to study the limits of applicability of 
the finite-size scaling law.}
Expansions have been revised with the help of the online 
Taylor Series Calculator of WolframAlpha. 
R. Garcia-Millan has enjoyed  a stay at the Centre de Recerca Matem\`atica
through its Internship Program, 
as well as a scholarship from the London Mathematical Laboratory.
The rest of authors acknowledge support from projects
FIS2012-31324, from Spanish MINECO, and 2014SGR-1307, from AGAUR.

\section{Appendix}

A fractional linear function is defined by
\begin{equation}
f(s) = \frac {a+bs}{c+ds},
\label{flf}
\end{equation}
with $a,b,c$ and $d$ constants fulfilling 
$a d \ne bc$
(to avoid that the numerator and the denominator are proportional).
Note that for the geometric distribution, Eq. (\ref{fgeo}), 
$a=p, b=0, c=1$, and $d=-q$,
although in the next paragraphs we will keep generality.

The advantge of fractional linear functions is that their compositions 
are very manageable.
To see this 
we follow the calculation of Karlin and Taylor
\cite{Karlin_Taylor}.
Let us consider any point $s_i$, then,
it is direct to see that
\begin{equation}
f(s)-f(s_i) = \left(\frac {cb-ad}{c+d s}\right)
\left(\frac {s-s_i}{c+d s_i}\right),
\end{equation}
and for two points $s_0$ and $s_1$ one has
\begin{equation}
\frac{f(s)-f(s_0)}{f(s)-f(s_1)}=\left(\frac{c+ds_1}{c+ds_0}\right)
\left(\frac{s-s_0}{s-s_1}\right).
\end{equation}
For fractional linear functions representing
probability generating functions 
there exist just two fixed points that, by definition, verify 
$s_i=f(s_i)$, so one can identify the previous $s_0$ and $s_1$
with these fixed points. 
It can be also verified that it is only at the critical point ($m=1$)
that the two fixed points take the same value, $s_0=s_1$. 
Note that the fixed point $s^*$ corresponding
to the probability of extinction in the infinite system (mentioned in Sec. 2)
is defined as $s^*=\min(s_0,s_1)$.
So, using the defining property of fixed points ($s_i=f(s_i)$) and defining
$\kappa=({c+ds_1})/({c+ds_0})$ and $w=f(s)$ one gets
\begin{equation}
\frac{w-s_0}{w-s_1}=\kappa \left(\frac{s-s_0}{s-s_1}\right).
\end{equation}
In order to calculate $f(w)$ one can iterate the same argument 
for the left-hand side of the equation, and in general,
by induction,
\begin{equation}
\frac{w_t-s_0}{w_t-s_1}=\kappa^t \left(\frac{s-s_0}{s-s_1}\right),
\end{equation}
with $w_t=f^t(s)$.
Isolating $w_t$ one arrives at the desired formula
for the compositions of $f(s)$,
\begin{equation}
w_t=f^t(s)=\frac{s_0-\kappa^t s_1(s-s_0)/(s-s_1)}{1-\kappa^t(s-s_0)/(s-s_1)},
\label{ladelapendice}
\end{equation}
{which holds for any values of the parameters of the offspring distribution, 
except at the critical point ($m=1$).}

In the case in which $f(s)$ is a probability generating function,
one of the fixed points is equal to one, 
by normalization. 
So, one can take, without loss of generality 
$s_1=1$. 
Substituting the form of a fractional linear function, 
Eq. (\ref{flf}), into $f(1)=1$ one gets a relation between
the parameters $a,b,c$ and $d$.
One can also verify that the other fixed point is $s_0=-a/d$.
Finally, the constant $\kappa$ turns out to be, substituting $s_0$,
\begin{equation}
\kappa=\frac{c+d} {c+d s_0} = \frac{c+d}{c-a},
\end{equation}
which happens to be identical to the inverse of the mean, 
i.e., \begin{equation}\kappa=m^{-1}.\end{equation}

For the generalized geometric distribution, 
from its probability generating function, Eq. (\ref{ggeopgf}),
and from the definition of fractional linear functions, Eq. (\ref{flf}), 
one establishes 
that $a=p_0$, $b=p-p_0$, $c=1$, and $d=-q$, 
and the fixed point $s_0$ turns out to be
\begin{equation}
s_0=\frac{p_0}q,
\end{equation} 
which, for the particular case of the geometric distribution,
defined by $p_0=p$, turns into
\begin{equation}
s_0=\frac{p}q.
\end{equation} 
The knowledge of the value of the fixed point $s_0$ leads 
to the explicit form for $f^t(s)$.

{At the critical point, given by $m=1$, it is necessary to follow a separate approach.
For the generalized geometric distribution, the critical point is given by 
$p=1-p_0$, which, substituting into the probability generating function, Eq. (\ref{ggeopgf}),
leads to
\begin{equation}
f(s)=\frac{p_0+(1-2p_0)s}{1-p_0s}.
\end{equation}
Induction leads directly to
\begin{equation}
f^t(s)=
\frac{(1-p_0)\{t p_0+[1-(t +1)p_0]s\}}{1+(t-2)p_0-(t-1)p_0^2 -(t p_0-t p_0^2) s }
\end{equation}
\begin{equation}
=\frac{t p_0+[1-(t +1)p_0]s}{1+(t-1)p_0-t p_0s},
\end{equation}
from where the order parameter of the transition turns out to be
\begin{equation}\rho(L)=1-P_{ext}(L) = 1-f^L(0) \simeq \frac{1-p_0}{L p_0}=
\frac 2 {\sigma_c^2 L},\end{equation}
taking the limit of large $L$ and using the expression
above for $\sigma_c^2$.
This is in perfect agreement with the results obtained for $m\ne 1$.
Note that the results for the geometric distribution are a particular case corresponding to
$p_0=p=1/2$ at $m=1$.
}


\begin{thebibliography}{10}

\bibitem{Barber}
M.~N. Barber.
\newblock Finite-size scaling.
\newblock In C.~Domb and J.L. Lebowitz, editors, {\em Phase Transitions and
  Critical Phenomena, Vol. 8}, pages 145--266. Academic Press, London, 1983.

\bibitem{Kuzemsky}
A.~L. Kuzemsky.
\newblock Thermodynamic limit in statistical physics.
\newblock {\em Int. J. Mod. Phys. B}, 28:1430004, 2014.

\bibitem{Bustamante}
C.~Bustamante, J.~Liphardt, and F.~Ritort.
\newblock The nonequilibrium thermodynamics of small systems.
\newblock {\em Phys. Today}, 58(7):43--48, 2005.

\bibitem{Parrondo}
J.~M.~R. Parrondo, J.~M. Horowitz, and T.~Sagawa.
\newblock Thermodynamics of information.
\newblock {\em Nature Phys.}, 11:131--139, 2015.

\bibitem{Brezin}
E.~Br\'ezin.
\newblock An investigation of finite size scaling.
\newblock {\em J. Phys.}, 43:15--22, 1982.

\bibitem{Privman}
V.~Privman.
\newblock Finite-size scaling theory.
\newblock In V.~Privman, editor, {\em Finite Size Scaling and Numerical
  Simulation of Statistical Systems}, pages 1--98. World Scientific, Singapore,
  1990.

\bibitem{Stanley}
H.~E. Stanley.
\newblock {\em Introduction to Phase Transitions and Critical Phenomena}.
\newblock Oxford University Press, Oxford, 1973.

\bibitem{Yeomans1992}
J.~M. Yeomans.
\newblock {\em Statistical Mechanics of Phase Transitions}.
\newblock Oxford University Press, New York, 1992.

\bibitem{Christensen_Moloney}
K.~Christensen and N.~R. Moloney.
\newblock {\em Complexity and Criticality}.
\newblock Imperial College Press, London, 2005.

\bibitem{Wolfram_Vieta}
E.~W. Weisstein.
\newblock Vieta's substitution.
\newblock {\em Mathworld}.
\newblock http://mathworld.wolfram.com/VietasSubstitution.html.

\bibitem{Galton_Watson}
H.~W. Watson and F.~Galton.
\newblock On the probability of the extinction of families.
\newblock {\em J. Anthropol. Inst. Great Britain Ireland}, 4:138--144, 1875.

\bibitem{Harris_original}
T.~E. Harris.
\newblock {\em The Theory of Branching Processes}.
\newblock Springer, Berlin, 1963.

\bibitem{branching_biology}
M.~Kimmel and D.~E. Axelrod.
\newblock {\em Branching Processes in Biology}.
\newblock Springer-Verlag, New York, 2002.

\bibitem{neutron_fluctuations}
I.~P\'azsit and L.~P\'al.
\newblock {\em Neutron Fluctuations}.
\newblock Elsevier, Oxford, 2008.

\bibitem{Corral_FontClos}
A.~Corral and F.~Font-Clos.
\newblock Criticality and self-organization in branching processes: application
  to natural hazards.
\newblock In M.~Aschwanden, editor, {\em Self-Organized Criticality Systems},
  pages 183--228. Open Academic Press, Berlin, 2013.

\bibitem{Alstrom}
P.~Alstr{\o}m.
\newblock Mean-field exponents for self-organized critical phenomena.
\newblock 38:4905--4906, 1988.

\bibitem{Zapperi_branching}
S.~Zapperi, K.~B. Lauritsen, and H.~E. Stanley.
\newblock Self-organized branching processes: Mean-field theory for avalanches.
\newblock {\em Phys. Rev. Lett.}, 75:4071--4074, 1995.

\bibitem{Pruessner_book}
G.~Pruessner.
\newblock {\em Self-Organised Criticality: Theory, Models and
  Characterisation}.
\newblock Cambridge University Press, Cambridge, 2012.

\bibitem{GarciaMillan}
R.~Garcia-Millan, F.~Font-Clos, and A.~Corral.
\newblock Finite-size scaling of survival probability in branching processes.
\newblock {\em Phys. Rev. E}, 91:042122, 2015.

\bibitem{Karlin_Taylor}
S.~Karlin and H.~M. Taylor.
\newblock {\em A first course in stochastic processes}.
\newblock Academic Press, San Diego, 2nd edition, 1975.

\bibitem{Harris52}
T.~E. Harris.
\newblock First passage and recurrence distributions.
\newblock {\em Trans. Am. Math. Soc.}, 73:471--486, 1952.

\bibitem{Stanley_rmp}
H.~E. Stanley.
\newblock Scaling, universality, and renormalization: {Three} pillars of modern
  critical phenomena.
\newblock {\em Rev. Mod. Phys.}, 71:S358--S366, 1999.

\end{thebibliography}
\end{document}